\documentclass[aps,prd,twocolumn,showpacs,floatfix,preprintnumbers,nofootinbib,superscriptaddress]{revtex4-1}

\usepackage{graphicx}
\usepackage{color}
\usepackage{natbib}
\usepackage{multirow}

\usepackage{epsfig}
\usepackage{graphicx}
\usepackage{amsfonts}
\usepackage{amssymb}
\usepackage{latexsym}
\usepackage{amsmath}
\usepackage{hyperref}

\def\ee{\end{equation}}
\def\ba{\begin{eqnarray}}
\def\ea{\end{eqnarray}}

\def\bq{\begin{quote}}
\def\eq{\end{quote}}

 at 10truept

\newcommand{\beq}{\begin{equation}}
\newcommand{\eeq}{\end{equation}}
\newcommand{\beqa}{\begin{eqnarray}}
\newcommand{\eeqa}{\end{eqnarray}}
\newcommand{\bea}{\begin{eqnarray}}
\newcommand{\eea}{\end{eqnarray}}

\def\lesssim{~\mbox{\raisebox{-.6ex}{$\stackrel{<}{\sim}$}}~}

\def\ltap{\ \raise.3ex\hbox{$<$\kern-.75em\lower1ex\hbox{$\sim$}}\ }
\def\gtap{\ \raise.3ex\hbox{$>$\kern-.75em\lower1ex\hbox{$\sim$}}\ }
\def\gl{\ \raise.5ex\hbox{$>$}\kern-.8em\lower.5ex\hbox{$<$}\ }
\def\roughly#1{\raise.3ex\hbox{$#1$\kern-.75em\lower1ex\hbox{$\sim$}}}

\begin{document}

\title{Phase Transitions in the Early Universe: Impacts on BBN and CMB Observables} 
\author{Rui Xu}
\affiliation{Department of Physics and Chongqing Key Laboratory for Strongly Coupled Physics, Chongqing University, Chongqing 401331, P. R. China}

\author{Jiachen Lu} 
\affiliation{Department of Physics and Chongqing Key Laboratory for Strongly Coupled Physics, Chongqing University, Chongqing 401331, P. R. China}

\author{Shihao Deng}
\affiliation{Department of Physics and Chongqing Key Laboratory for Strongly Coupled Physics, Chongqing University, Chongqing 401331, P. R. China}

\author{Ligong Bian}
\email{lgbycl@cqu.edu.cn}
\affiliation{Department of Physics and Chongqing Key Laboratory for Strongly Coupled Physics, Chongqing University, Chongqing 401331, P. R. China}

\begin{abstract}

We explore the effects of low-scale cosmological first-order phase transitions on Big Bang Nucleosynthesis (BBN) and Cosmic Microwave Background (CMB) anisotropy and distortion. We examine two scenarios: the distribution of the phase-transition energy in the cosmic background and the phase transition as an additional source of energy injection. Our analysis reveals that the CMB spectrum from the Planck 2018 dataset imposes stringent constraints on sub-MeV-scale phase transitions, while keV-scale phase transitions can induce significant CMB distortions. Consequently, BBN and CMB observables serve as complementary tools to constrain phase-transition parameters, alongside gravitational wave detections.

 \end{abstract}

\maketitle

\section{Introduction}

Cosmological phase transitions (PTs) have quite a few effects on the evolution of the universe, and these effects are basically produced by first-order PTs~\cite{Hindmarsh:2020hop}, like baryogenesis~\cite{Hall:2019ank}, magnetic fields~\cite{Guth:1979bh, ellis2019intergalactic}. 
First-order PTs also have the potential to generate gravitational waves (GWs). The release of latent heat and associated bubble collisions or turbulence during PT can produce a stochastic background of primordial GWs~\cite{Bian_2021, Wang:2020jrd, Hindmarsh:2017gnf, Caprini:2015zlo, Caprini:2019egz, Di:2020ivg, Jinno:2021ury, Kamionkowski:1993fg}. These GWs carry information about early-universe physics beyond the Standard Model~\cite{Bian:2021ini, Cai:2017cbj, Caldwell:2022qsj}, and could be detected by LISA~\cite{Baker:2019nia, LISA:2017pwj}, Taiji~\cite{Ruan:2018tsw}, LIGO~\cite{aasi2015advanced} and pulsar timing arrays (PTAs)~\cite{NANOGrav:2020bcs, Goncharov:2021oub, EPTA:2021crs}.

BBN is highly sensitive to the universe’s expansion rate, as it determines the abundances of primordial elements like deuterium~\cite{Pitrou:2020etk}, helium~\cite{Pitrou:2018cgg}, and lithium~\cite{depta2021acropolis}. 
First-order PTs can modify the expansion history by altering the energy density and the equation of state. If such a transition occurs near or before BBN, it can impact element formation~\cite{Bai:2021ibt}, allowing BBN observations to constrain PTs properties. CMB anisotropies are an important phenomenon in observation, energy injection close to the recombination affects CMB anisotropy~\cite{Paoletti_2019, Niedermann_2020}. Additionally, CMB spectral distortions~\cite{Chluba:2018cww, Lucca_2020, Liu:2023fgu, Liu:2023nct, Chluba_2014, Chluba_2011} that diverge the spectrum from the black body spectrum can observe changes in the energy distribution of photons.
Previous studies have focused on first-order PTs in a dark sector with specific triggering mechanisms to address the Hubble tension~\cite{Niedermann_2020, Niedermann_2021, Hayashi_2023, garny2024hotnewearlydark}, here we focus on how does the first-order PT modify the energy evolution in the early Universe and affect the CMB spectrum. 

In this paper, we investigate the effect of the first-order phase transition of sub-MeV on BBN and CMB and constrain the PT parameters with observational data, including the PT time, strength, and duration. We first briefly introduce the PTs dynamic in section II, and discuss the BBN and CMB predictions from the low-scale slow first-order PTs in section III and V, the conclusion is drawn in the last section.

\section{Phase transition}
 
During the first-order PT, after a critical temperature $T_c$, nucleation of bubbles of the new phase begins, and these bubbles grow and expand, eventually transforming the entire space into the new phase. The nucleation rate of the phase transition can be described as~\cite{PhysRevD.15.2929, enqvist1992nucleation}:  
\begin{equation}
    \begin{aligned}
&\Gamma(t)=\Gamma_0 e^{\beta t}\;,
    \end{aligned}
\end{equation} 
where $\beta$ determines the characteristic timescale of the phase transition, with the duration estimated as $\Delta t \sim \beta^{-1}$. In radiation dominated era, the pre-factor can be approximated as 
\begin{equation}
    \Gamma_0^{1 / 4} = \left(\frac{4 \pi^3 g_{\star}}{45} \right)^{1/2} \left(\frac{T_p^2}{m_{\mathrm{Pl}}} \right) e^{-\beta / 8 H_{\star}},
\end{equation}  
where \(T_p\) is the temperature at which the phase transition occurs, and \(H_{\star}\) represents the corresponding Hubble parameter. The Planck mass is given by \(m_{\mathrm{Pl}} \approx 1.22 \times 10^{19} ~\mathrm{GeV}\)~\cite{Planck:2018vyg}.  

During the first-order PT, the evolution of the universe is governed by the Friedmann and Boltzmann equations~\cite{Liu:2022lvz, Liu:2021svg}  
\begin{equation}
    H^2 = \frac{1}{3} \left( \rho_{\gamma} + \rho_w + \rho_v \right)\;,
\end{equation}  

\begin{equation}
    \frac{d}{dt} \left( \rho_{\gamma} + \rho_w \right) + 4 H \left( \rho_{\gamma} + \rho_w \right) = - \frac{d \rho_v}{dt}\;.
    \label{eqtransf}
\end{equation}  
Here, the false vacuum energy density is expressed as \(\rho_v = F(t) \Delta V\), with \(F(t)\) denoting the probability of the field remaining in the false vacuum state, and \(\Delta V\) being the energy difference between the false and true vacuum. The quantities \(\rho_{\gamma}\) and \(\rho_w\) represent the energy densities of radiation and bubble walls, respectively. The evolution of vacuum energy follows  
\begin{equation}
    -\frac{d \rho_v}{dt} = -\alpha \rho_\gamma \cdot \frac{dF(t)}{dt},
\end{equation}  
where the dimensionless parameter \(\alpha = \Delta V / \rho_{\gamma}\) characterizes the strength of the phase transition.  

The radius of a bubble nucleated at time \(t'\) is given by: 
\begin{equation}
    r\left(t, t^{\prime}\right) = \int_{t^{\prime}}^t a^{-1}(t) dt,
\end{equation}  
where \(a(t)\) is the scale factor. The function \(F(t)\) follows~\cite{turner1992bubble}  
\begin{equation}
    F(t) = \exp \left[-\frac{4 \pi}{3} \int_{t_i}^t dt' \Gamma\left(t'\right) a^3\left(t'\right) r^3\left(t, t'\right) \right],
\end{equation}  
where \(t_i\) is the onset time of the phase transition. Initially, the field remains in the false vacuum with \(F(t < t') = 1\). We specifically consider the case where the average survival probability at a given time \(t_p\) is \(F(t_p) = 0.7\).  

\begin{figure}[!htp]
\begin{center}
  \includegraphics[width=0.45\textwidth]{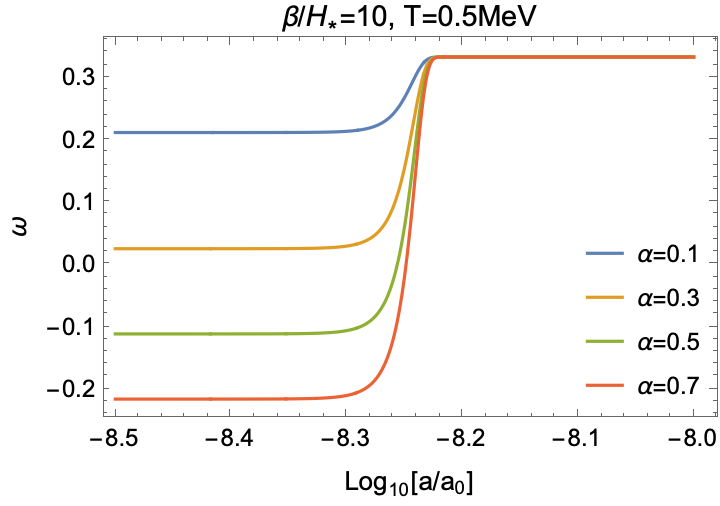}
  \includegraphics[width=0.45\textwidth]{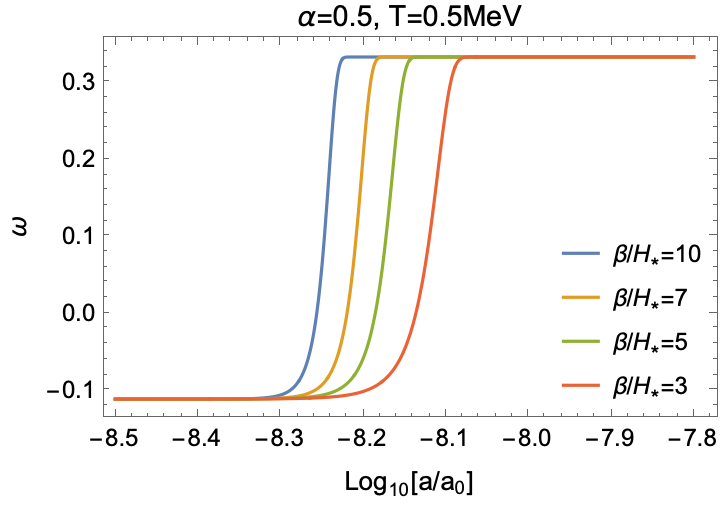}
  \caption{The equation of state parameter $\omega$ versus the the scale factor $a$ for different PT strengths and inverse durations, $a_0$ is today's value.}
  \label{ptomega}
  \end{center}
\end{figure}

The PT which redistributes energy from \(\rho_v\) to \(\rho_\gamma\) can alter the Hubble rate. We consider the total energy density \(\rho = \rho_\gamma + \rho_v\), which scales as \(\rho \propto a^{-3(1+\omega)}\). Figure~\ref{ptomega} illustrates the changes in the equation of state \(\omega\) for different PT parameters.

\section{Big Bang Nucleosynthesis and PTs}   

\begin{figure}[!htp]
\begin{center}
\includegraphics[width=0.45\textwidth]{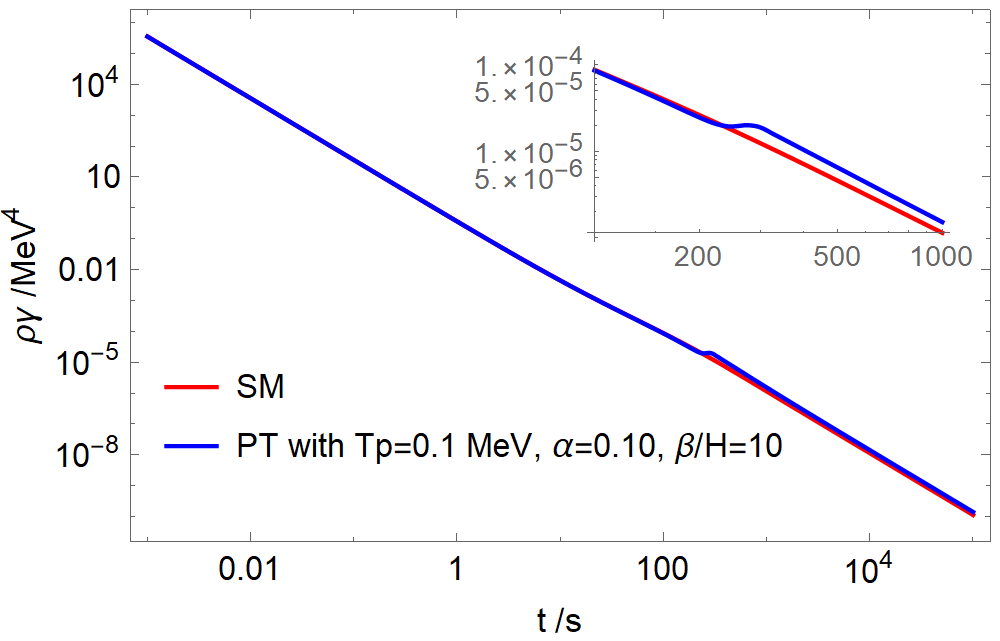}

		\caption{The evolution of photon energy density over time. The red line corresponds to the standard model, while the blue line represents the case with a PT, where the PT parameter is  $T_p =0.1\,\text{MeV}$, $\alpha = 0.1$, and $\beta/H_{*} = 10$.}
		\label{fig:rhogammat}
		\end{center}
	\end{figure}

To demonstrate the PT's effect on the neutrino decoupling process when $T_p$ is below $1\,\text{MeV}$, we present the case of photon reheating. In the Fig.~\ref{fig:rhogammat}, we show the evolution of photon energy density over time, illustrating the specific impact of the first-order PT on cosmic thermodynamics. In the case of photon reheating, it can be observed that during the PT, the photon energy density increases, thereby affecting neutrino decoupling.

 \begin{figure}[!htp]
\begin{center}
\includegraphics[width=0.45\textwidth]{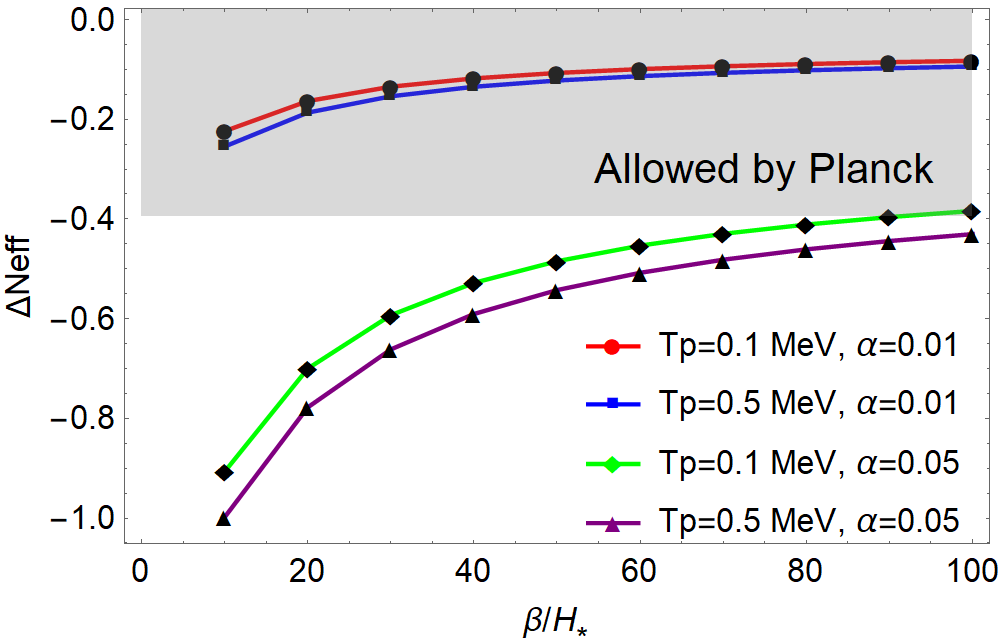}
  \caption{The evolution of $\Delta N_{\text {eff }}=N_{\text {eff }}-N_{\text {eff }}^{\mathrm{SM}}$ (with $N_{\mathrm{eff}}^{\mathrm{SM}}=3.044(1)$~\cite{EscuderoAbenza:2020cmq, Akita:2020szl, Froustey:2020mcq, Bennett:2020zkv})  over $\beta/H_{*}$. The upper two lines illustrate the variation of the value of $\Delta N_{\rm eff}$  with $\beta/H_{*}$ under the conditions $\alpha = 0.01$. And the lower two  lines corresponds to $\alpha = 0.05$. The gray region shows the Planck 95\% confidence level (CL) constraints 
 $N_{\text {eff }}= 2.99^{+0.33}_{-0.34}$~\cite{Planck:2018vyg}.}
  \label{fig:DNEFF}
  \end{center}
 \end{figure}

\begin{figure}[!htp]
\begin{center}
\includegraphics[width=0.45\textwidth]{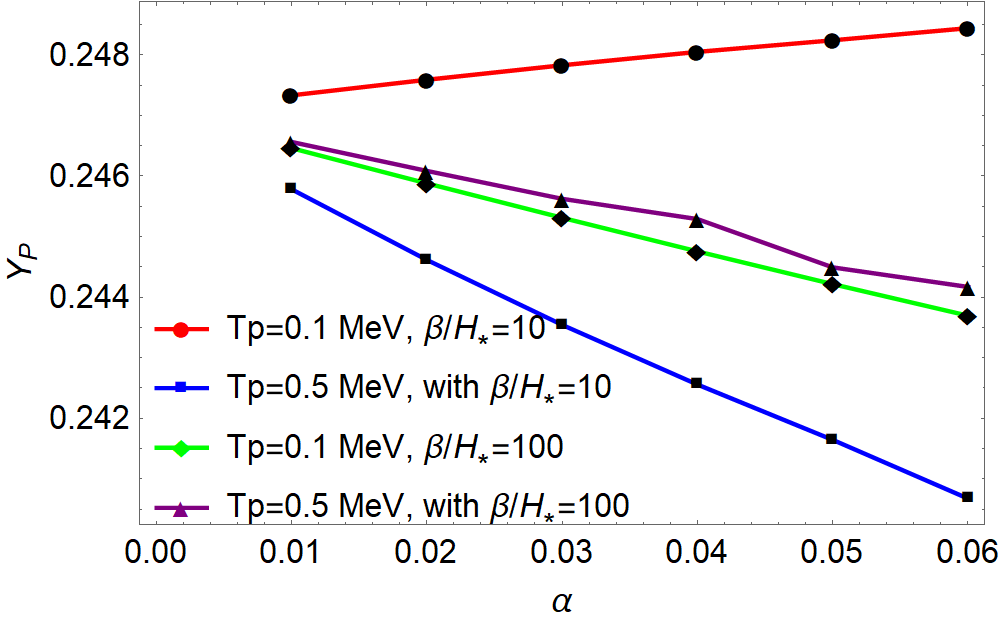}
\includegraphics[width=0.45\textwidth]{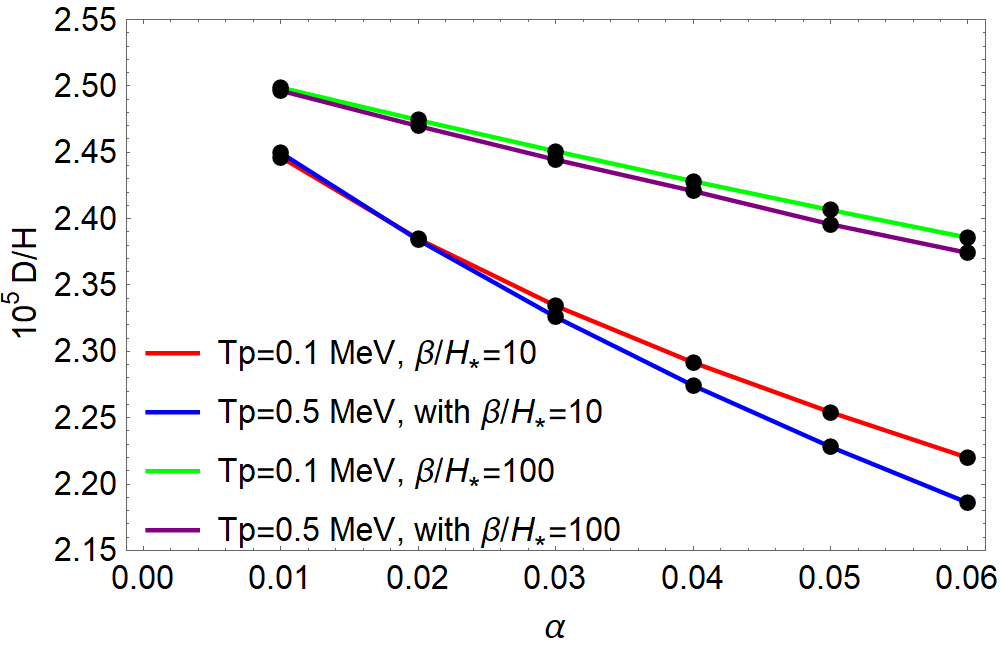}

		\caption{ The helium and deuterium abundances versus the PT strength $\alpha$ for different $T_p$ and $\beta/H_\star$}
  \label{fig:YPDHA}
		\end{center}
		\end{figure}
 

Fig.~\ref{fig:DNEFF} mainly illustrates the impact of $\beta/H_{*}$ on $\Delta N_{\rm eff}$, while also reflecting the effects of $\alpha$ and $T_p$. The upper two lines correspond to $\alpha= 0.01$, while the lower two corresponds to $\alpha= 0.05$. The red and green lines represent $T_p=0.1\,\text{MeV}$, and the blue and purple lines represent $T_p=0.5\,\text{MeV}$. It can be observed that as $\alpha$ increases and $\beta/H_{*}$ decreases, the impact of the PT becomes more significant, leading to a smaller $N_{\rm eff}$. 

We considered the impact of PTs on BBN ~\cite{Deng:2023twb}. In the early universe, protons and neutrons were in thermal equilibrium. As the temperature decreased, the neutron fraction gradually declined. When the temperature reached approximately 0.078 MeV, the deuterium abundance peaked, and helium began to be synthesized in large quantities.
The helium abundance is mainly determined by $X_{n}(t_{\rm FO})$ and $t_{\rm nuc}$. Considering the effects of PTs, changes in these parameters alter $X_n(t_{\rm nuc})$, which directly impacts the final helium abundance, approximated as $Y_{\rm P}\approx 2X_n(t_{\rm nuc})$. For deuterium, its final abundance is primarily influenced by the timing of its destruction. Photon (neutrino) reheating from first-order PTs modifies the time-temperature relation, extending (shortening) the deuterium destruction period, resulting in a lower (higher) final ${\rm D/H}|_{\rm P}$.

 Fig.~\ref{fig:YPDHA} illustrates $\alpha$ and $T_p$ on helium and deuterium abundances. It can be observed that as $\alpha$ increases, both $Y_{\rm P}$ and ${\rm D/H}|_{\rm P}$ decrease for $\beta/H_{*} = 100$, while for $\beta/H_{*} = 10$, helium abundance $Y_{\rm P}$ increases, which comes from a subtle effect, as discussed in ~\cite{Deng:2023twb}.
 Specifically, for $\beta/H_{*} = 100$, the influence of $t_{\rm nuc}$ dominates over that of $T_{\rm FO}$, whereas for $T_p =0.1\,\text{MeV}$, $\beta/H_{*}=10$, the opposite is true.


\section{CMB with first-order PT}

In this section, we consider the effects of the PT on background energy transformation and therefore the CMB anisotropy, and also the CMB distortion driven by the additional energy injection  from the PT.

\subsection{Background energy distribution}

The helium abundance $Y_{\rm P}$ plays a crucial role in determining the optical depth   
\begin{equation}
\kappa\left(\eta_L\right) \equiv \int_{t_L}^{t_0} \frac{d t}{\tau(t)}=\int_{\eta_L}^{\eta_0} \sigma_T n_t X_e a(\eta) d \eta.
\end{equation}
leading to greater suppression of CMB anisotropies, particularly at small scales, where $n_t=\frac{3 H_0^2 \Omega_b}{8 \pi G m_H} (1-Y_{\rm P})$ is today total electron number, $\sigma_T$ is Thomson scattering cross-section, $X_e=n_e/n_H$ is  free electron fraction, $\eta$ is conformal time. Visibility function

\begin{equation}
g=\frac{d \kappa} {d \eta} \cdot e^{-\kappa}=\sigma_T n_t X_e \cdot a \cdot e^{-\kappa},
\end{equation}
where \( a \) is the scale factor. The visibility function reaches its maximum at the recombination epoch, corresponding to the scale factor \( a_L \).


We utilize the publicly available cosmological code CLASS~\cite{lesgourgues2011, Diego_Blas_2011} to compute CMB anisotropy power spectra, the
optical depth $\kappa$ enters into the source function for CMB anisotropy~\cite{Lesgourgues_2014}
\begin{equation}
S_{T 0}=g\left(\frac{1}{4} \delta_\gamma+\psi\right)+e^{-\kappa}\left(\phi^{\prime}+\psi^{\prime}\right),
\end{equation}
where $\frac{1}{4} \delta_\gamma$ is intrinsic temperature fluctuation, $\psi$ is gravitational redshift term, $\phi$ arises from gauge perturbations. The source function \(S_{T 0}\) is included in the photon transfer function:
\begin{equation}
\Delta_{\ell}^{T_j}(q)=\int_{\tau_{\mathrm{ini}}}^{\tau_0} d \tau S_{T_j}(k, \tau) \phi_{\ell}^{j}(\nu, \chi),
\end{equation}
where $\phi_{\ell}^{j}$ are radial functions, this can be used to calculate correlation function
\begin{equation}
C_{\ell}^{TT}=4 \pi \int \frac{d k}{k} \Delta_{\ell}^{T}\left(q, \tau_0\right) \Delta_{\ell}^{T}\left(q, \tau_0\right) \mathcal{P}(k),
\end{equation}
where $\mathcal{P}(k)$ is the primordial spectrum, $q$ is a function of $k$.
Here, we choose two benchmarks with PT temperatures being 0.5 MeV and 0.1 MeV to investigate how does the PT impact on the CMB. 

\begin{figure}[!htp]
\begin{center}
    \includegraphics[width=0.45\textwidth]{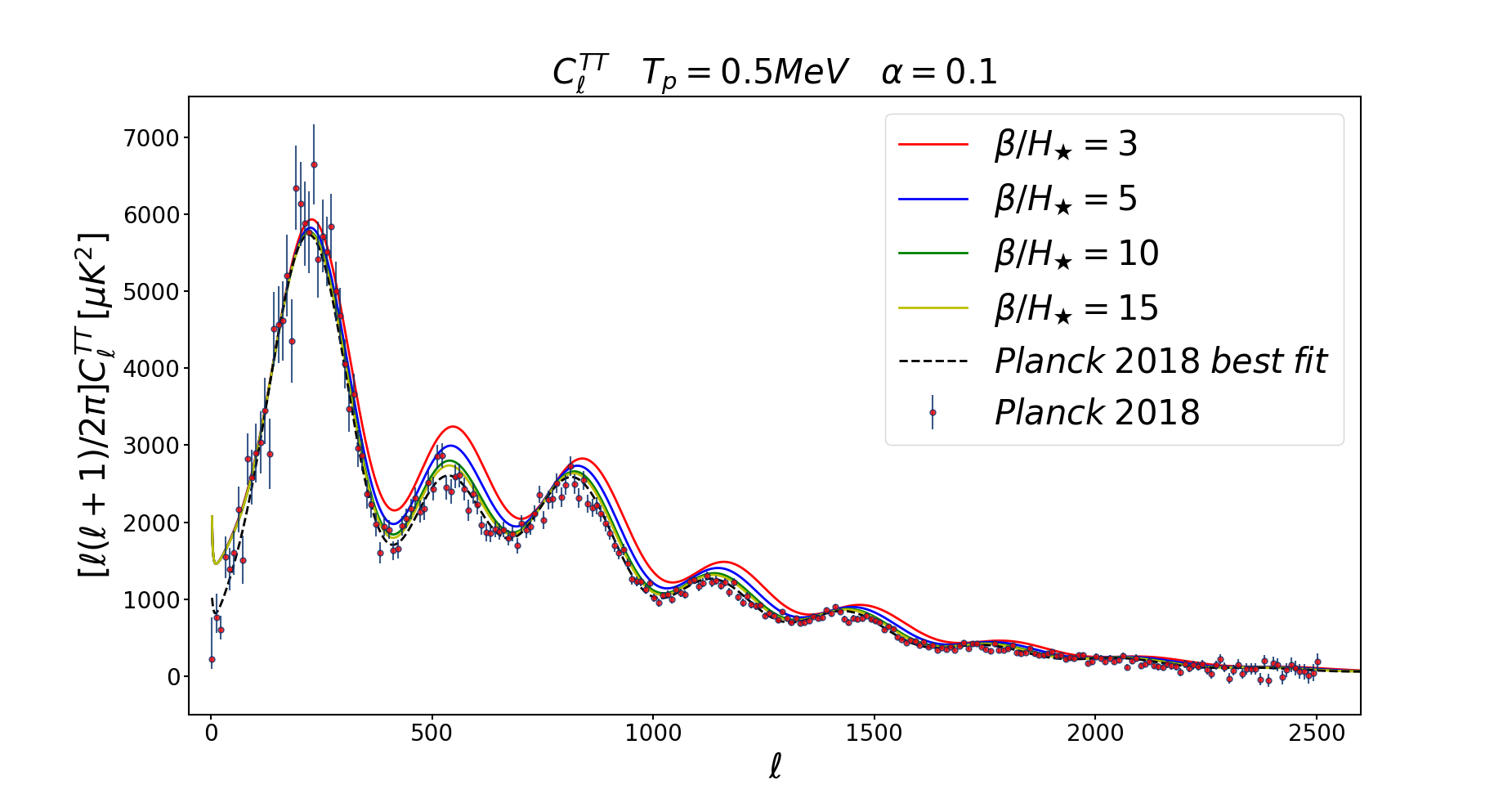}
    \includegraphics[width=0.45\textwidth]{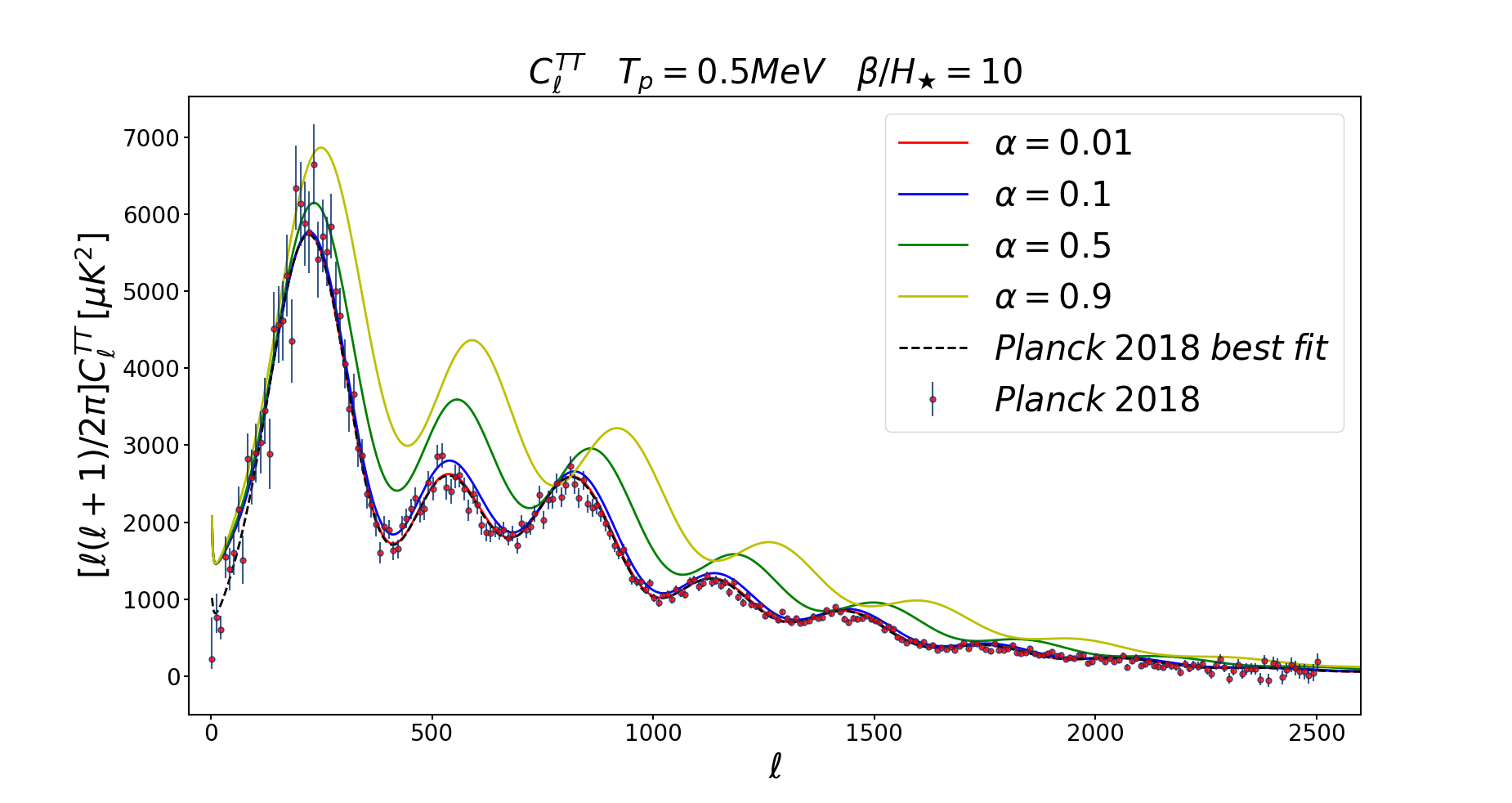}
\caption{The effects PT parameters on the CMB TT spectrum, with the PT temperature being 0.5 MeV. The vertical lines with red dots represent the error bars from the Planck 2018 data~\cite{Planck:2018vyg}.}
\label{cmb_05MeV}
\end{center}
\end{figure}

Figure~\ref{cmb_05MeV} displays the angular power spectrum of correlation functions for temperature \(T\) anisotropies, with each curve representing different configurations of the PT parameters. Variations in the parameter \(\alpha\) strongly affect the angular power spectrum, particularly around the peak, where large changes are observed. The Planck 2018 data is more restrictive on \(\alpha\), with values typically around 0.1. The data also allow us to set a strong limit on the duration, requiring \(\beta/H_{\star} \gtrsim 5\). 

\begin{figure}[!htp]
\begin{center}
    \includegraphics[width=0.45\textwidth]{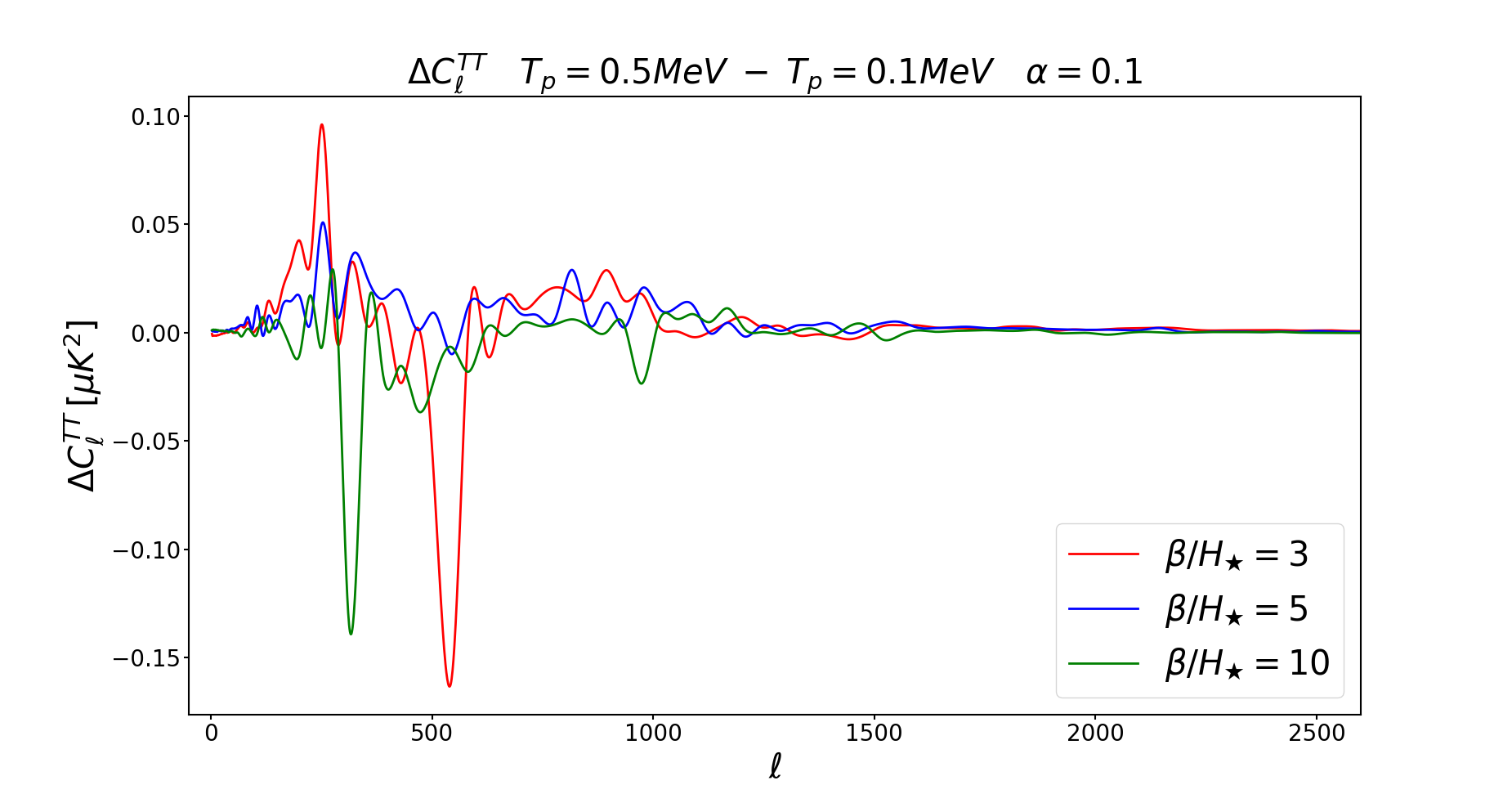}  
    \includegraphics[width=0.45\textwidth]{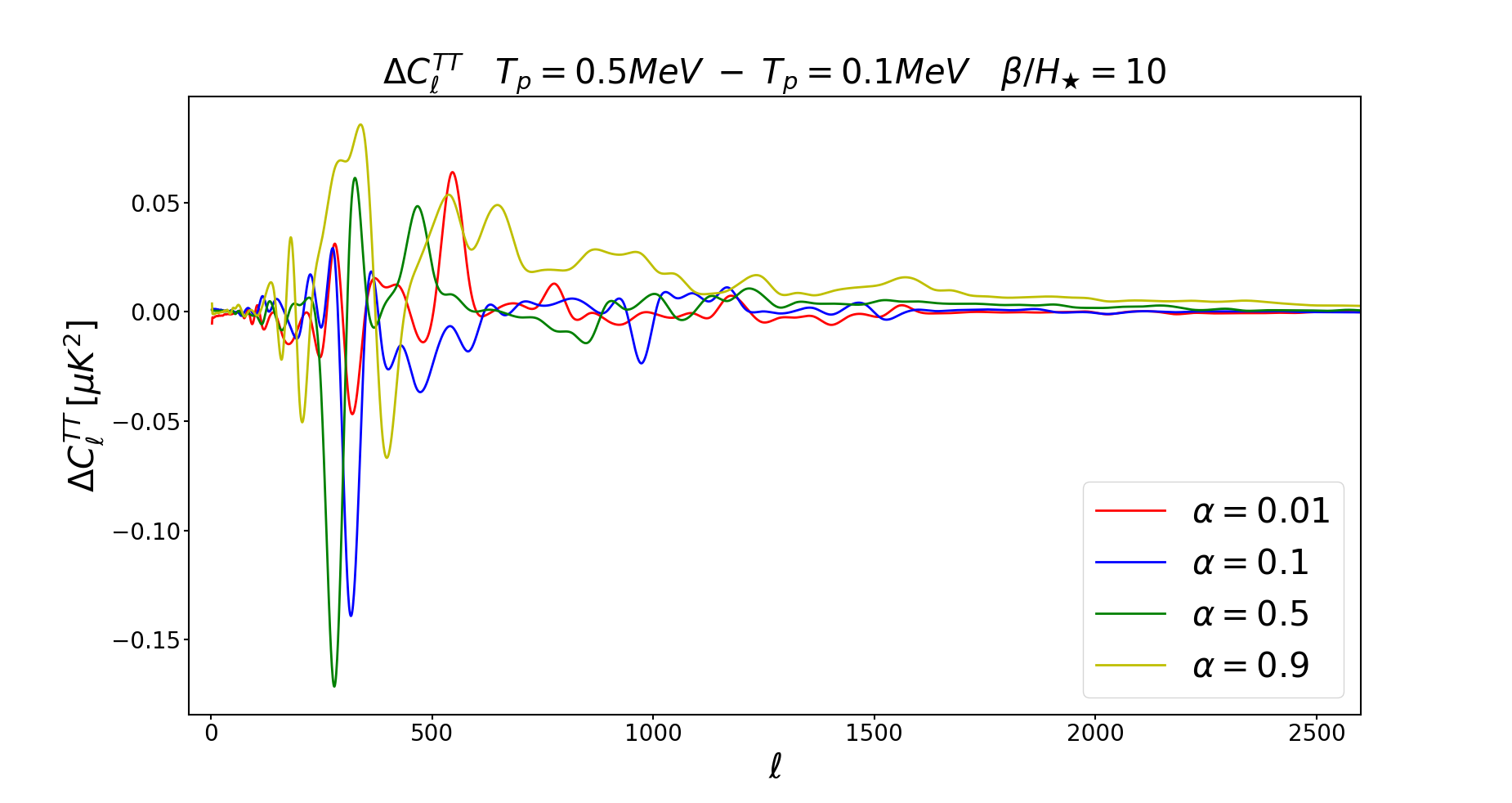}   
 \caption{The difference between the CMB anisotropy spectra for PTs at 0.5 MeV and 0.1 MeV.}
\label{cmb_differ}
\end{center}
\end{figure}

\begin{figure}[!htp]
\begin{center}
    \includegraphics[width=0.45\textwidth]{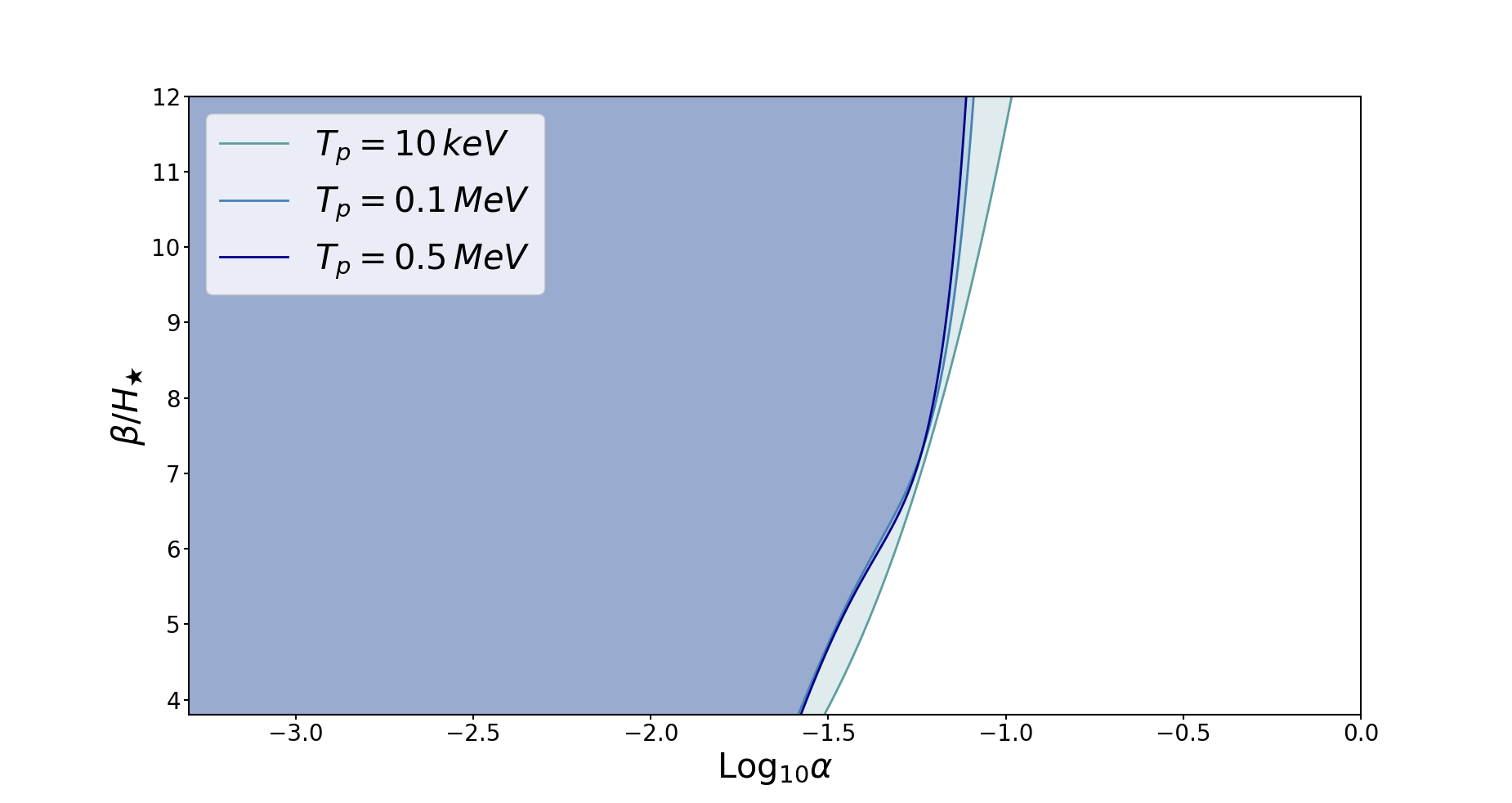}  
 \caption{The shaded regions represent the allowed parameter spaces for different PT temperatures by Planck 2018 data.}
\label{PT_region}
\end{center}
\end{figure}

In Fig.~\ref{cmb_differ}, we show the difference between the two CMB power spectra to highlight these subtle variations. The spectra for the PT at 0.5 MeV and 0.1 MeV do not look very different because the PT occurs far from the CMB epochs, and after the vacuum energy is converted into radiation energy, it evolves proportionally to $a^{-4}$. 
In Fig.~\ref{PT_region}, we further restrict the parameter space of PTs with the Planck 2018 data. 
The plot highlights the constraints on the $\beta/H_{\star}$ and $\alpha$ weakly depends on the PT temperature, and PT parameter spaces with $\alpha\gtrsim\mathcal{O}(10^{-1})$ have been excluded for the slow PT at sub-MeV scales.

\begin{figure}[!htp]
\begin{center}  
    \includegraphics[width=0.5\textwidth]{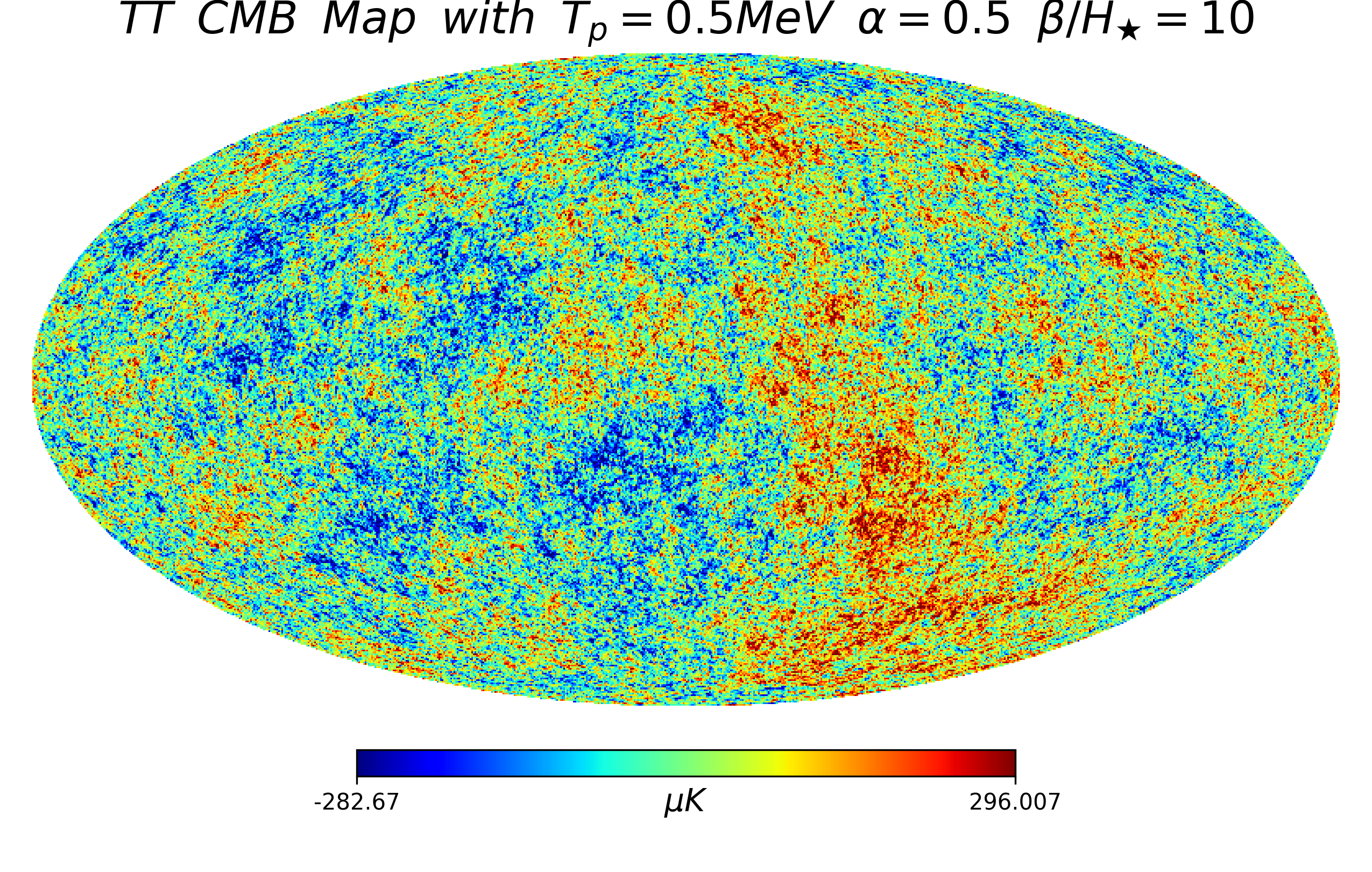}    
\caption{CMB temperature map with PT Effects.}
\label{fig:SKY}
\end{center}
\end{figure}

Figure~\ref{fig:SKY} presents the CMB temperature map, incorporating the effects of a first-order PT. The color scale represents temperature fluctuations relative to the mean CMB temperature (2.7255K), where red regions indicate slightly higher temperatures and blue regions indicate lower temperatures.

Our analysis employs a phase transition mechanism that differs from that of New Early Dark Energy (NEDE)~\cite{Niedermann_2020}. While NEDE models the PT as a source of early dark energy, we focus on a different scenario where vacuum energy is converted into radiation after the phase transition. In NEDE, the PT occurs very rapidly, characterized by \(\beta/H_{\star} > 100\), with a corresponding redshift of \(z_* \sim 5000\). In contrast, our PT is significantly slower, with \(\beta/H_{\star} \sim 10\), and occurs at a much higher redshift of approximately \(z \sim 10^9\), close to the epoch of BBN. In our approach, CMB observational data constrain the parameters \(\alpha\) and \(\beta/H_{\star}\). These constraints ensure that the PT remains weak and occurs over a very short duration, effectively approximating an instantaneous transition. Our constraints suggest that a prolonged or strong transition would introduce deviations in the CMB anisotropy spectra, which are inconsistent with current observational data.

\subsection{PT with CMB distortion}
In the previous section, we discussed the limitations imposed on PTs occurring near the epoch of BBN. Next, we will discuss PTs occurring between BBN and the CMB, which can be constrained by spectral distortions. There are essentially two types of sources for spectral distortions: non-injection and injection. Non-injection distortions do not involve direct energy or particle injection. Instead, they can be caused by processes such as variations in the equation of state during the PTs. Non-injection distortions sourced by first-order PTs are too small to be detected. Therefore, we now focus on distortions generated by PTs as a source of energy injection.
 
At redshifts \(z > 2 \times 10^6\), the evolution of photon number and energy density is highly efficient, maintaining the CMB spectrum as a blackbody. However, at lower redshifts \(z < 2 \times 10^6\), the processes related to changes in photon number become less efficient, leading to the onset of spectral distortions. This period is characterized by efficient Compton scattering, which generates a chemical potential and is referred to as the \(\mu\)-era. Following the \(\mu\)-era, the \(y\)-era begins when Compton scattering also becomes inefficient, resulting in \(y\)-type spectral distortions. This era spans from \(z = 5 \times 10^4\) to the present. The \(\mu\)-era  begin corresponds to a cosmic temperature of approximately 500 eV. Spectral distortions are generated when energy injection occurs within this temperature range. Therefore, in this analysis, we focus on PTs at the KeV scale, which are close to the \(\mu\)-era.

\begin{figure}[!htp]
\begin{center}
  \includegraphics[width=0.4\textwidth]{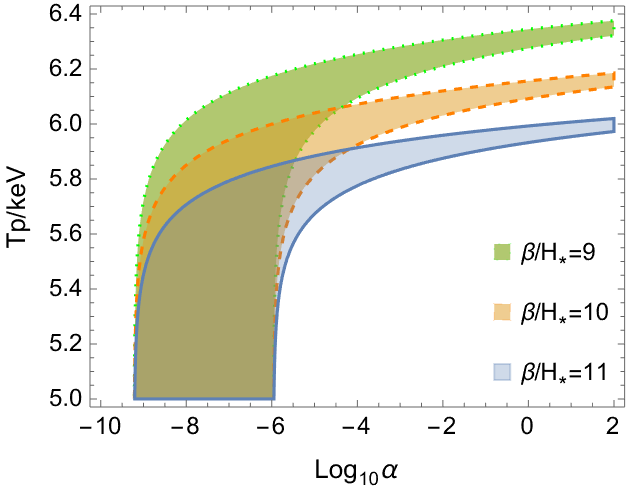}
  \caption{The $\mu$ distortion versus different PT parameters.}
  \label{distortion_inj}
  \end{center}
\end{figure}

The heating rate $\dot{Q}=\frac{\partial \rho}{\partial t}+4 H \rho$ ~\cite{Lucca_2020}, with Eq.~\ref{eqtransf}  we get the energy injection from PT
\begin{equation}
    \begin{aligned}
&\dot{Q}_{p t}=-\frac{d \rho_v}{d t}=-\alpha \rho_\gamma \cdot \frac{d F(t)}{d t} .
    \end{aligned}
\end{equation}
The $\mu$ distortion from the first-order PT is
\begin{equation}
    \begin{aligned}
&\mu=1.4 \int_{z_1}^{z_2} d z \frac{\frac{d Q_{pt}}{d z}}{\rho_\gamma} e^{-\left(\frac{z}{z_{D C}}\right)^{\frac{5}{2}}},
    \end{aligned}
\end{equation}
where $z_1=2 \times 10^6$, $z_2=5 \times 10^4$ , and
\begin{equation}
    \begin{aligned}
&z_{D C} \equiv 1.97 \times 10^6\left[1-\frac{1}{2}\left(\frac{Y_{\rm P}}{0.24}\right)\right]^{-\frac{5}{2}}\left(\frac{\Omega_b h^2}{0.0224}\right)^{-\frac{2}{5}}\;.\nonumber
    \end{aligned}
\end{equation}
The transformation between redshift \(z\) and time \(t\) is given by \(dt = -\frac{dz}{(1+z) H(z)}\). During radiation domination, time increases as \(t \propto a^2\). The \(\mathrm{COBE/FIRAS}\) experiment~\cite{Fixsen_1996} sets a limit of \(|\mu| < 9 \times 10^{-5}\), while \(\mathrm{PIXIE}\)~\cite{A_Kogut_2011} sets a more stringent limit of \(|\mu| < 5 \times 10^{-8}\). In Figure~\ref{distortion_inj}, we show the range \(5 \times 10^{-8} \sim 9 \times 10^{-5}\) for \(\mu\) distortion, which lies between the limits of the two detectors. To the bottom-right of these curves, the PT regions with large $\alpha$ and small $T_p$ are excluded by the \(\mu\) distortion. As the strength of the PTs increases, the PT must occur earlier to produce the same level of spectral distortion. This behavior contrasts with the dependence on the parameter \(\beta / H_{\star}\), where a increase in \(\beta / H_{\star}\) requires the PT to occur later.

\section{Gravitational Wave}

During the first-order PT, expanding bubbles of the new phase collide and transfer energy to the surrounding plasma, creating motion in the form of sound waves. The GW spectrum of the sound waves is characterized by the strength of the PT $\alpha$, the inverse duration 
 $\beta/H$, the PT temperature, and the bubble wall velocity. We use GW spectrum given in Ref.~\cite{NANOGrav:2020bcs, Hindmarsh_2015, Guo_2021, Tamanini_2016, Huber_2008}:
\begin{eqnarray}
    \Omega_{\mathrm{GW}}^{\mathrm{sw}} h^2(f)&=&7.7 \times 10^{-5} g_*^{-\frac{1}{3}}\left(\frac{\kappa_{\mathrm{sw}} \alpha_*}{1+\alpha_*}\right)^2\left(\frac{H_*}{\beta}\right) 0.513 v_w \nonumber\\
&& \times\left(1-\left(1+2 \tau_{\mathrm{sw}} H_*\right)^{-\frac{1}{2}}\right)\left(f / f_p^{\mathrm{sw}}\right)^3\nonumber\\
&&\times\left[7 /\left(4+3\left(f / f_p^{\mathrm{sw}}\right)^2\right)\right]^{\frac{7}{2}}\;,
\end{eqnarray} 
where $f_p^{\mathrm{sw}}=1.13 \times 10^{-10} \mathrm{~Hz} \frac{0.536}{v_w}\left(\frac{\beta}{H_*}\right)\left(\frac{T_*}{\mathrm{MeV}}\right)\left(\frac{g_*}{10}\right)^{1 / 6}$, $v_w$ is bubble wall velocity, $\kappa_{sw}$ is fraction of vacuum energy transform into sound wave, $\tau_{sw}H_{\star}$ is about lifetime of sound wave. 

\begin{figure}[!htp]
\begin{center}
 \includegraphics[width=0.45\textwidth]{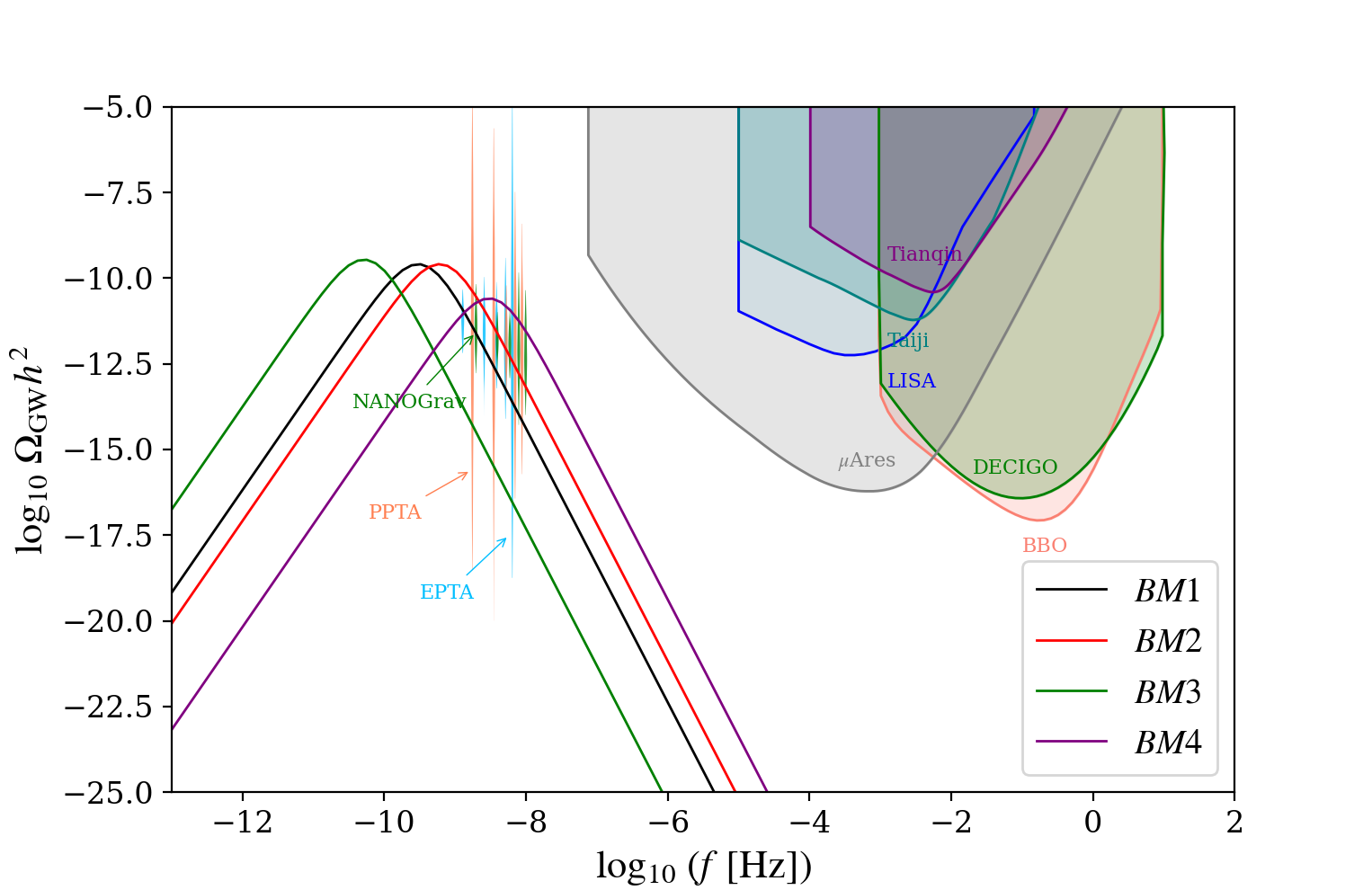}
 \caption{We show the results from the four benchmark points listed in Table~\ref{tabgw}. Solid lines show GW spectra from sound wave with different parameter choice, where $v_w$=1, $\tau_{sw}H_{\star}=1, \kappa_{sw}=0.2$. We also present the sensitivities of different GW detectors, i.e. $\mu$Ares~\cite{Sesana:2019vho} LISA~\cite{Baker:2019nia,LISA:2017pwj}, Taiji~\cite{Ruan:2018tsw, Hu:2017mde}, Tianqin~\cite{TianQin:2015yph},  DECIGO~\cite{Seto:2001qf, Kawamura:2020pcg, Kudoh:2005as}, BBO~\cite{Crowder:2005nr} and include NANOGrav~\cite{NANOGrav:2020bcs}, PPTA~\cite{Goncharov:2021oub} and EPTA~\cite{EPTA:2021crs}.}
\label{sw}
\end{center}
\end{figure}

As illustrated in the Figure~\ref{sw}, the resulting GW signals from the PTs with $0.1 \lesssim\alpha $, $10 \lesssim\beta/ H_{\star} \lesssim 100$ at sub-MeV scales 0.1 MeV $\lesssim T_{\star}\lesssim$ 1 MeV fall within the sensitivity range of PTAs' observations. Therefore, the sub-MeV scale slow first-order PT can be complementarily constrained by GW detectors and CMB anisotropy spectra.

\begin{table}[ht]
\centering 
\caption{Benchmark in Fig~\ref{sw}.} 
\label{tabgw}
\begin{tabular}{cccc}
\hline \hline 
 & $\alpha$ & $\beta/H_{\star}$ & $T_{\star}$(MeV) \\
\hline
$BM_1$ & 0.1 & 10 & 0.5  \\
$BM_2$ & 0.1 & 10 & 1  \\
$BM_3$ & 0.1 & 10 & 0.1  \\
$BM_4$ & 0.1 & 100 & 0.5  \\
\hline \hline
\end{tabular}
\end{table}

\section{Conclusions}
In this work, we consider possibility of a low-scale first-order PT occurring close to the time of BBN and after BBN. With the observation that: 1) the $\Delta N_{\rm eff}$ has a very strong restriction on the strength and durations of the PT; 2) both the values of $Y_P$ and $D/H|_P$ generally decrease as the PT strength increases, and the tendency is more significant for slower PTs. The PT induced energy evolution in the background together with the BBN yields the influence on the CMB anisotropy. We found that the CMB anisotropy strongly restrict the strength of the PT as $\alpha\lesssim 0.1$ and the inverse duration of the PT as $\beta/H_{\star}\gtrsim 5$. And, the effect of the PT temperature is very weak, on the order of $\mathcal{O}(10^{-1})$. When the PT occurs around KeV scale, we observe that the CMB distortion from energy injection, with the effect of different PT inverse durations converging to unity at very small $\alpha$, and a higher PT temperature is accompanied with a slower PT. The sensitivity of PTAs to the detection of GWs limits the strong PT with $\alpha\gtrsim \mathcal{O}(10^{-1})$ for slow PTs with $5\lesssim\beta/H_{\star}\lesssim 10$.

\begin{acknowledgments}
This work is supported by the National Key Research and Development Program of China under Grant No. 2021YFC2203004, and by the National Natural Science Foundation of China (NSFC) under Grants Nos. 12322505, 12347101. We also acknowledges Chongqing Talents: Exceptional Young Talents Project No. cstc2024ycjh-bgzxm0020 and Chongqing Natural Science Foundation under Grant No. CSTB2024NSCQ-JQX0022.

\end{acknowledgments}

\bibliography{ptbbn}

\end{document}